\documentclass[conference]{IEEEtran}

\usepackage{graphicx} % Required for the inclusion of images
\usepackage{amsmath} % Required for some math elements
\usepackage{fancyhdr}
\usepackage{indentfirst}
\usepackage{float}
\usepackage{setspace}
\usepackage{subfigure}
\usepackage{stfloats}
\usepackage{subfigure}
\usepackage{dsfont}
\usepackage{cite}
\usepackage{comment}
\usepackage{amsmath}
\usepackage{algorithm}
\usepackage{algorithmic}
\usepackage{amssymb}
\begin{document}
\title{Experimental Study on Reference-Path-Aided System Calibration for mmWave Bistatic ISAC Systems}

\author{\IEEEauthorblockN{Chenhao Luo\IEEEauthorrefmark{1}, Chongrui Wang\IEEEauthorrefmark{1} Aimin Tang\IEEEauthorrefmark{1}, Fei Gao\IEEEauthorrefmark{2}, Chaojun Xu \IEEEauthorrefmark{2}}
\IEEEauthorblockA{\IEEEauthorrefmark{1} Global College, Shanghai Jiao Tong University,
Shanghai, China\\
\IEEEauthorrefmark{2} Noika Bell Labs China, Shanghai, China\\
Email: {\{chenhao.luo, chongrui.wang, tangaiming\}@sjtu.edu.cn,\{fei.e.gao, chaojun.xu\}@nokia-sbell.com}}}

\maketitle

\begin{abstract}
%\boldmath

Integrated sensing and communications (ISAC) has been regarded as a key enabling technology for next-generation wireless networks. Compared to monostatic ISAC, bistatic ISAC can eliminate the critical challenge of self-interference cancellation and is well compatible with the existing network infrastructures. However, the synchronization between the transmitter and the sensing receiver becomes a crucial problem. The extracted channel state information (CSI) for sensing under communication synchronization contains different types of system errors, such as the sampling time offset (STO), carrier frequency offset (CFO), and random phase shift, which can severely degrade sensing performance or even render sensing infeasible. To address this problem, a reference-path-aided system calibration scheme is designed for mmWave bistatic ISAC systems, where the line-of-sight (LoS) path can be blocked. By exploiting the delay-angle sparsity feature in mmWave ISAC systems, the reference path, which can be either a LoS or a non-LoS (NLoS) path, is first identified. By leveraging the fact that all the paths suffer the same system errors, the channel parameter extracted from the reference path is utilized to compensate for the system errors in all other paths. A mmWave ISAC system is developed to validate our design. Experimental results demonstrate that the proposed scheme can support precise estimation of Doppler shift and delay, maintaining time-synchronization errors within 1 nanosecond.
\end{abstract}

\IEEEpeerreviewmaketitle

\section{Introduction}
Integrated sensing and communication (ISAC) has emerged as a key technology in next-generation wireless networks \cite{survey}. By jointly designing the communication and sensing functionalities, ISAC systems offer several advantages over conventional architectures, including improved spectral efficiency, hardware reuse, and mutual enhancement between communication and sensing performance \cite{survey2}.

There are two basic modes to integrate sensing function with communication networks: monotonic ISAC and bistatic ISAC. In monostatic ISAC, the transmitter and sensing receiver are co-located or integrated in the same transceiver. Therefore, the full-duplex radio serves as the key enabler, with self-interference cancellation being the primary challenge \cite{tang2024interference}. In bistatic ISAC, the transmitter and sensing receiver are separate devices and usually deployed at different positions, e.g., they can be two different base stations (BS). Half-duplex radios can be utilized to support bisatic sensing. Given that existing communication network infrastructures are based entirely on half-duplex radios, bistatic sensing emerges as a promising approach for broad deployment in future ISAC networks. Since sensing and communication share the same physical channel in bistatic ISAC systems\cite{Luo2024Channel}, the communication receiver's hardware can be fully reused for wireless sensing. For example, the wireless sensing can be achieved by processing the received OFDM symbols or channel state information (CSI) \cite{OFDMsensing}. However, since the transmitter and receiver are separate devices, the synchronization between the two devices becomes a key challenge for wireless sensing \cite{wu2024sensing}.

A communication receiver usually utilizes pilots or reference signals to synchronize with the transmitter and estimates the CSI. However, the accuracy of synchronization in communications is not enough to support accurate sensing. Therefore, when utilizing the estimated CSI for wireless sensing, additional calibration is mandatory for sampling timing offset (STO), carrier frequency offset (CFO), and random phase shift. The STO introduces a bias in range estimation, whereas the CFO and random phase shift cause time-varying system errors, rendering Doppler estimation and coherent combination infeasible. So far, several approaches have been developed for system calibration in bistatic ISAC systems.
%In \cite{tewes2021ws}, the transmitter and receiver are connected via a wired link, allowing them to share a common clock and thereby avoid synchronization impairments. However, In bistatic ISAC systems, the ISAC transmitter and ISAC receiver are physically separated, making direct access to shared reference clocks or wired synchronization links infeasible. Consequently, over-the-air (OTA) calibration becomes essential in calibrating bistatic ISAC systems. Several researches have studied OTA calibration in bistatic ISAC systems.

If the system errors are consistent over all antennas, one specific antenna can be utilized as a reference to calibrate the time-varying system errors. One typical method is the CSI-ratio scheme, where the ratio of the CSI on two antennas is calculated to compensate for time-varying system errors \cite{zeng2019farsense}. Another typical method is the cross-antenna cross-correlation (CACC) \cite{qian2018widar2}, where the conjugate multiplication is performed between the CSI of the remaining antennas and that of the reference antenna. Additional steps are required to remove the side effects of the CSI-ratio and CACC, such as virtual paths in CACC \cite{wang2022single}.
%reference, such as the length of the line-of-sight (LoS) path, is still required to calibrate STO.
%In \cite{qian2018widar2}, a specific antenna is designated as the reference antenna, and the conjugate multiplication is performed between the CSI of the remaining antennas and that of the reference antenna to eliminate the unknown timing offset and carrier frequency offset. In \cite{zeng2019farsense}, the ratio of the CSI on two antennas is calculated to compensate for time- and frequency-dependent, yet spatially independent, system errors.
%However, both approaches rely on reserving one antenna exclusively for calibration purposes. As a result, the CSI from the reference antenna cannot be exploited for sensing tasks, thereby limiting the utilization of spatial diversity and potentially degrading the overall sensing performance.
%Apart from leveraging the consistency over antennas or in the spatial domain, the consistency of the static strong path is also utilized for calibration.
%In \cite{li2023integrating}, a strong line-of-sight (LoS) path is set, and its channel parameter is extracted to compensate for the phase offsets caused by both STO, CFO, and random phase shift. % Due to the strong signal power and temporal stability, the LoS path is assumed to exhibit a consistent channel response over time. The estimated channel is then divided by the distorted channel impulse response (CIR) of the LoS path to cancel the phase distortion and sampling time offsets in the system.
To calibrate both the STO and time-varying system errors, delay-domain features can be extracted, as all paths experience the same system impairments. In \cite{li2023integrating}, a strong line-of-sight (LoS) path is established as a reference, and its channel parameters are used to compensate for the phase offsets induced by the STO, CFO, and random phase shift. In \cite{meneghello2022sharp}, the compressed sensing algorithm is developed to extract the LoS path information due to the low resolution in the delay domain.
%Similarly in \cite{samczynski20215g}, the signal from the direct link between the transmitter and receiver is utilized as the reference. The cross-ambiguity function is then obtained by calculating the cross-correlations between the signal of the LoS path and other paths to cancel the STO.
%Besides, other studies directly estimate system errors and then conduct compensation.
In \cite{zhao2023multiple}, the static paths and dynamic paths are first separated, and then the random phase extracted from the static paths is used to compensate for the Doppler estimation of dynamic paths. In \cite{pegoraro2024jump}, calibration for the mmWave IEEE 802.11ay system is considered, where the LoS path is identified as the first peak in the channel impulse response (CIR), which is then used to compensate for STO and cancel CFO with the static feature of the LoS path. However, these methods rely on the LoS path and assume the same STO for all antennas.

In this paper, a reference-path-aided system calibration scheme is designed for mmWave bistatic OFDM ISAC systems. %The sensing receiver fully reuses the OFDM communication hardware to extract CSI for sensing. the extracted CSI contains different STOs for different receive antennas.
Moreover, a more general system model is considered, where different antennas experience distinct STOs, and the LoS path may be blocked. To handle the blockage of the LoS path, an additional reference reflector, such as a reconfigurable intelligent surface (RIS), is deployed to create a non-LoS (NLoS) reference path for calibration.
%Thus, the raw CSI extracted from the OFDM symbols cannot be directly utilized for AoA estimation. In addition to the CFO, there exists a random phase shift along with time in the raw CSI, which renders Doppler estimation infeasible.
%To calibrate system errors, we also utilize a static and strong reference path, which can be a LoS or non-LoS (NLoS) path. %If there exists a LoS path, the LoS path can be used as the reference path. Otherwise,
To conduct system calibration, we first identify the reference path via the delay-angle sparsity feature in mmWave systems. Particularly, by leveraging the coherent delays over different antennas, the relative STOs over different antennas are first eliminated, and thus the AoA of different paths with different delays can be estimated. The reference path is further identified with the AoA information. Next, the channel parameter of the reference path is extracted. Since all the paths suffer the same system errors, the extracted channel parameter of the reference path is utilized to compensate for the system errors in all the paths via a simple ratio calculation. However, the ratio calculation will break the steering vector of each path, so the reconstruction of a phased array is finally conducted.
To validate the above design, we develop a mmWave OFDM ISAC system at 26 GHz with 500 MHz bandwidth.
The effectiveness of the proposed system calibration is validated by experimental studies under both LoS and NLoS scenarios, which shows that the random phase shift over time can be well eliminated for Doppler estimation, and a sub-nanosecond time synchronization can be achieved for range estimation.

%The rest of this paper is organized as follows. The system and signal models are presented in Section \uppercase\expandafter{\romannumeral2}. The detailed reference-path-aided calibration design is elaborated in Section \uppercase\expandafter{\romannumeral3}. Experiment validation of our developed calibration scheme is carried out in \uppercase\expandafter{\romannumeral4}. This paper is concluded in \uppercase\expandafter{\romannumeral5}.

%\input{calibration}
\section{System and Signal Models}

\subsection{System Model}
\begin{figure}[t]
    \centering
    \includegraphics[width=0.7\linewidth]{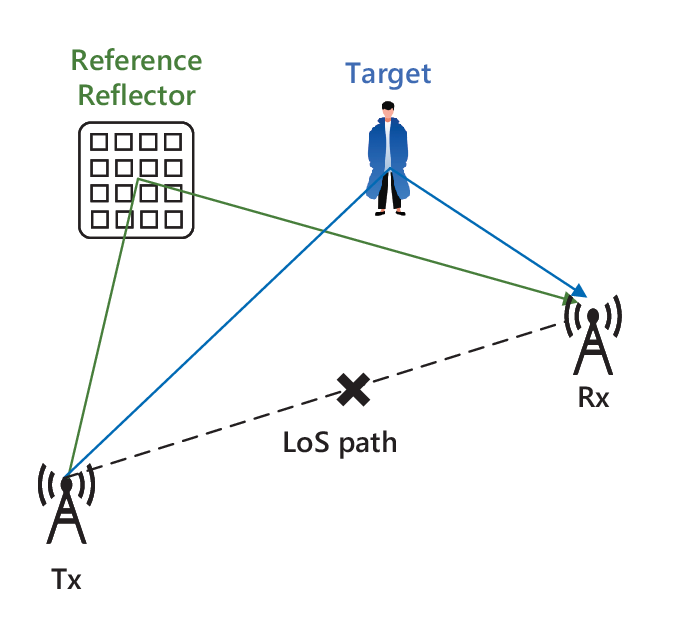}
    \vspace{-2em}
    \caption{Bistatic ISAC system model.}
    \vspace{-1em}
    \label{fig:sysmodel}
\end{figure}

In this paper, a bistatic ISAC system is considered, where the transmitter (Tx) and sensing receiver (Rx) localizations are fixed. For example, the Tx and sensing Rx can be two BSs in B5G/6G networks. In practice, the direct link, i.e., the LoS path between the Tx and Rx, may experience intermittent blockage due to moving obstacles, as shown in Fig.~\ref{fig:sysmodel}. %For example, the LoS path may be temporarily obstructed when a vehicle or train passes between the base stations, and it will then be restored once the obstruction clears.
Therefore, there may either be a LoS path between the transmitter and sensing receiver, or only NLoS paths between them. For the LoS scenario, the LoS path can be utilized as a reference path for system calibration. To support the NLoS scenario, a strong NLoS path is required for system calibration, which can be found in the environment or manually deployed. For instance, a RIS can be deployed as a reference reflector to form a strong NLoS path. The orientations and locations of the Tx, Rx, and reference reflector are prior knowledge. Therefore, in the LoS scenario, the angle of arrival (AoA) of the LoS path and the distance between the Tx and Rx are provided, whereas in the NLoS scenario, the AoA of the reference path and the combined distance along the Tx-reference reflector-Rx path are given.

%Consider the ISAC system shown in Fig. \ref{fig:sysmodel}. Depending on the location of Rx and the obstacles, the propagation conditions can be either the line-of-sight (LoS) case, as is shown in the link between Tx and Rx 1 or the NLoS case, as is shown in the link between Tx and Rx 2. Since both propagation scenarios are common, it is essential to consider both of them during system calibration. To conduct system calibration, we assume that some knowledge of the layout is known, including the location and orientation of Tx and Rx, and the location of reference reflectors. The reference reflectors are the  reconfigurable intelligent surfaces (RIS) placed in the environment to form a non-line-of-sight (NLoS) path between the Tx and Rx. Besides, the system is assumed to be a single-input-multiple-output (SIMO) system, where signals received at multiple Rx ports can be utilized.

% \begin{figure}[t!]
%     \centering
%     \subfigure[LoS scenario]{
%         \includegraphics[width=0.38\linewidth]{Fig/sysdiagLOS.pdf}
%         \label{fig:sysmodelLoS}
%     }
%     \subfigure[NLoS scenario]{
% 	    \includegraphics[width=0.55\linewidth]{Fig/sysdiagNLOS.pdf}
%         \label{fig:sysmodelNLoS}
%     }
%     \caption{Diagram of the two propagation conditions}
%     \label{fig:sysmodel}
% \end{figure}

\subsection{Signal Model}
The sensing Rx leverages the public reference signal (RS) embedded in the OFDM communication signal to extract the CSI for wireless sensing. Considering an OFDM signal of $N$ subcarriers and $M$ symbols, the received signal can be modeled as
%Consider a bistatic OFDM system of $N$ subcarriers with 1 transmit antenna at the transmitter and $P$ receive antennas at the receiver. Within the signal transmitted, reference signals (RS) are allocated for sensing purpose. The channel state information (CSI) is then obtained by utilizing these RS symbols. The calibration is conducted in each band individually. Within each band, the signal model is as follows. For the transmitted signal in the $k$-th band $\mathbf{X}_k$, the received signal $\mathbf{Y}_{k,p}$ at the $p$-th antenna in the $k$-th band is given by
\begin{equation}
    \mathbf{Y}_{p} =  \mathbf{H}_{p} \mathbf{X} +  \mathbf{w}_p,
    \label{eq:channel}
\end{equation}
where $\mathbf{X}$ is the transmitted signal, $\mathbf{H}_{p}$ is the CSI matrix at the $p$-th receive antenna, and $\mathbf{w}_p$ is the Gaussian noise. For the ideal case, the element of CSI matrix $\mathbf{H}_{p}$ on the $n$-th subcarrier of the $m$-th symbol can be modeled as
\begin{equation}
{\mathbf{H}}_{p}[m,n]=\sum_{l=1}^L{\boldsymbol{a}_{\boldsymbol{l}}[p]b_{l}}e^{-j2\pi n\Delta f \tau _l}e^{j2\pi m f_{\text{D},l}T},
\label{eq:oriCSI}
\end{equation}
where $L$ is the total number of propagation paths from the Tx to the sensing Rx; $\boldsymbol{a}_{\boldsymbol{l}}$ is the steering vector for the $l$-th path and $\boldsymbol{a}_{\boldsymbol{l}}=[1,e^{j2\pi\frac{d}{\lambda}\sin\phi_l},...,e^{j2\pi(P-1)\frac{d}{\lambda}\sin\phi_l}]$, where $d$ is the distance interval of adjacent antenna, $\lambda$ is the wavelength of the signal, and $\phi_l$ is the AoA of the $l$-th path; %$\boldsymbol{a}_{\boldsymbol{l}}[p]$ is the $p$-th element of the steering vector, which is the element at $p$-th antenna. The angle-of-arrival of the $l$-th path is denoted as $\phi_l$.
$\tau_l$ and $f_{\text{D},l}$ denote the delay and Doppler shift of the $l$-th path, respectively; $\Delta f$ and $T$ denote the OFDM subcarrier spacing and symbol duration, respectively;
$b_{l}$ is the complex attenuation of the $l$-th path. % For simplicity, we assume the subcarrier interval $\Delta f$ is identical in all $k$ bands. $f_{\text{D},l}$ denotes the Doppler shift of the $l$-th path, while $T$ is the time duration of each symbol.
The sensing estimation of delay $\tau _l$, doppler shift $f_{\text{D},l}$, and AoA $\phi_l$ mainly depends on the phase change over different dimensions.

Due to the hardware impairment between Tx and sensing Rx, there exist STO, CFO, and random phase shift in the extracted CSI at the sensing Rx. Without proper calibration of these errors, the sensing performance will be significantly influenced. %In bistatic systems, the signals are received at the receiver and channel estimations are then conducted. We apply the least square (LS) estimator to obtain the estimated channel state information $\hat{\mathbf{H}}$. Due to these system errors, the estimated channel will encounter additional error patterns compared with the ideal CSI in Eq. (\ref{eq:oriCSI}), which is given by
%To simplify the notation, we consider the CSI model in a single carrier/band and drop the index $k$.
Thus, considering these errors, the practical extracted raw CSI $\hat{\mathbf{H}_p}$ can be modeled as
\begin{equation}
\begin{aligned}
        \hat{\mathbf{H}}_{p}[m,n]=&e^{j\theta _m}e^{j2\pi n\Delta f \tau_{\mathrm{STO},p}}\sum_{l=1}^L{\boldsymbol{a}_{\boldsymbol{l}}[p]b_{l}}e^{-j2\pi n\Delta f \tau _l} e^{j\varphi_{\text{D},m,l}}.
    \label{eq:errCSI}
\end{aligned}
\end{equation}
The system errors in Eq. (\ref{eq:errCSI}) are explained as follows. Each antenna $p$ is connected with an independent RF chain, but they share the same clock. Therefore, the CFO and random phase shift along with time are the same for all antennas/RF chains. We use $\theta _m$ to represent the effect of both CFO and random phase shift over time. In the baseband, the samplings of different RF chains are handled independently, and thus the STO for different antennas may be different, which is denoted as $\tau_{\text{STO},p}$ for antenna $p$. Due to this issue, the AoA cannot be directly estimated with the raw CSI, even though the RF front-end is a phased array. $e^{j\varphi_{\text{D},m,l}} = e^{j2\pi m f_{\text{D},l}T}$ is used to denote the Doppler shift for simplicity. The extracted raw CSI is also under the impact of noise, which is ignored in Eq. (\ref{eq:errCSI}).
The main purpose of system calibration is to eliminate the impact of $\theta _m$ and $\tau_{\text{STO},p}$.

\section{Reference-Path-Aided Calibration Design}

\begin{figure}[t]
    \centering
    \includegraphics[width=\linewidth]{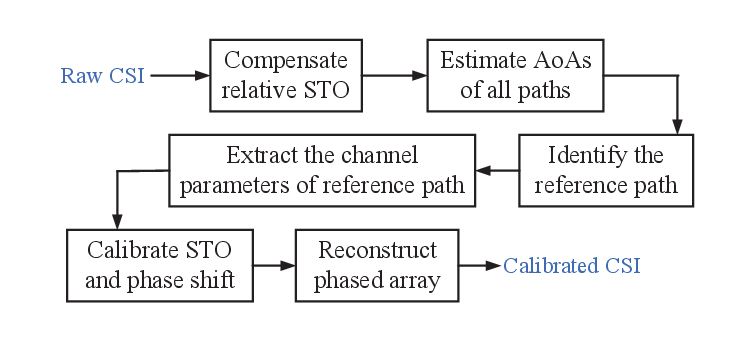}
     \vspace{-2em}
    \caption{Workflow of reference-path-aided calibration scheme.}
    \vspace{-1em}
    \label{fig:refcabflow}
\end{figure}

In the proposed system calibration scheme, a reference path is utilized to assist calibration. The reference path can either be the LoS path or an NLoS path, and thus this calibration scheme is feasible in both LoS and NLoS scenarios. There are certain requirements for the reference path. First, the reference path should be static and have a strong reflection power to make the path distinguishable and keep its channel response consistent.
%Besides, for the LoS reference path, the relative location of Tx and Rx should be known; for the NLoS reference path, apart from the location of Tx and Rx, the location of the reference reflector should also be known. The location information provides the essential parameters of the range and direction of the path, which is key in system calibration.

Unlike many existing studies that assume the presence of a LoS path easily identifiable as the first delayed path, our system model accounts for scenarios where the LoS path may be absent or intermittently blocked, requiring the use of a deployed NLoS reference for system calibration.
To this end, identifying the reference path with the raw CSI is the first problem to resolve. An intuitive solution is to check whether several key parameters of a path match the known information of the reference path. Although the range and AoA information of the reference path are given, they cannot be directly extracted from the raw CSI for matching. On the one hand, due to the existence of STO, the estimated delay/range of each path has a large bias, and thus cannot be matched with the exact range of the reference path. On the other hand, since different antennas have different $\tau_{\text{STO},p}$, the angle of each path cannot be accurately estimated.

To address this problem, we first eliminate the relative STOs for different antennas so that an accurate AoA estimation for each path can be enabled. After that, we utilize the estimated AoA information to identify the reference path. Next, the channel parameter of the reference path is extracted to compensate for the system errors in all other paths with a simple ratio calculation. However, the ratio calculation will break the steering vectors for each path, so a phased array reconstruction step is finally developed. The workflow of the proposed reference-path-aided calibration scheme is shown in Fig. \ref{fig:refcabflow}. More details are explained as follows.

\begin{figure}[t]
    \centering
    \subfigure[Original CIR]{
        \includegraphics[width=0.46\linewidth]{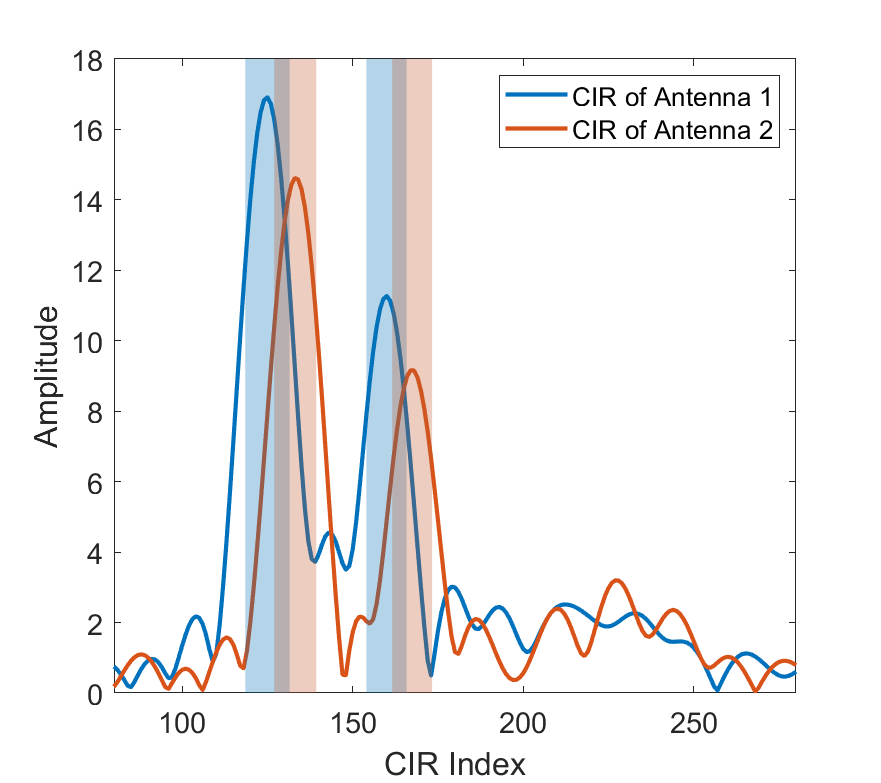}
    }
    \subfigure[Slid CIR of best match]{
	    \includegraphics[width=0.46\linewidth]{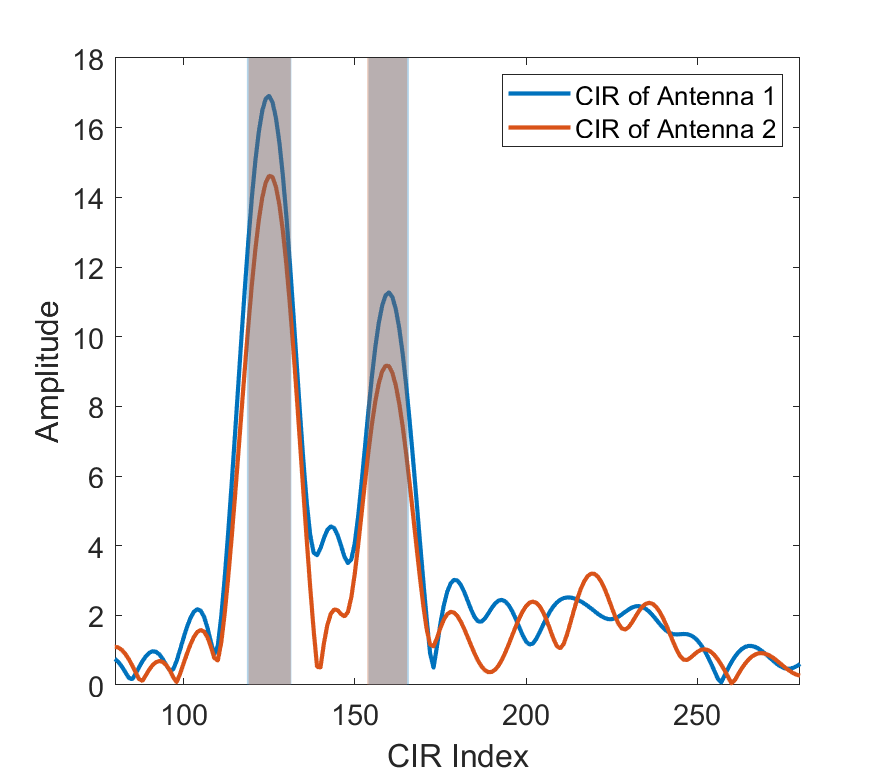}
    }
    \caption{Illustration of relative STO calibration.}
     \vspace{-1em}
    \label{fig:relSTO}
\end{figure}

\newcounter{TempEqCnt}                   % 创建临时变量TempEqCnt
\setcounter{TempEqCnt}{\value{equation}} % 将当前公式序号 赋给TempEqCnt
\setcounter{equation}{7}                 % 当前公式序号变为x，x等于长公式应有的序号减1.

\begin{figure*}[hb]
{\noindent} \rule[-10pt]{18cm}{0.05em}
    \begin{equation}
    \begin{aligned}
        \hat{\mathbf{H}}_{p,\mathrm{div}}[m,n]
        &=\tilde{\mathbf{H}}_{p}[m,n]e^{-j2\pi n\Delta f(\tau _{\mathrm{ref}}-u_{\mathrm{ref}}\delta )}/(\frac{\tilde{\mathbf{Q}}_{p}[m,u_{\mathrm{r}}]}{| \tilde{\mathbf{Q}}_{p}[m,u_{\mathrm{r}}] |})\\
        &=\frac{e^{j\theta _m}e^{j2\pi n\Delta f\tau _{\mathrm{STO},1}}(\boldsymbol{a}_{\mathbf{ref}}[p]b_{\mathrm{ref}}e^{-j2\pi n\Delta f\tau _{\mathrm{ref}}}+\sum_{\substack{l=1\\l\neq\mathrm{ref}}}^L{\boldsymbol{a}_{\boldsymbol{l}}[p]b_{l}}e^{-j2\pi n\Delta f\tau _l}e^{j\varphi_{\text{D},m,l}})e^{-j2\pi n\Delta f(\tau _{\mathrm{ref}}-u_{\mathrm{r}}\delta )}}{e^{j\theta _m}\boldsymbol{a}_{\mathbf{ref}}[p]}\\
    &=b_{\mathrm{ref}}e^{-j2\pi m\Delta f\tau _{\mathrm{ref}}}+\sum_{\substack{l=1\\l\neq\mathrm{ref}}}^L{(\boldsymbol{a}_{\boldsymbol{l}}[p]/\boldsymbol{a}_{\mathbf{ref}}[p])b_{l}}e^{-j2\pi n\Delta f\tau _l}e^{j\varphi_{\text{D},m,l}})\\
    \end{aligned}
    \label{eq:CSIdiv}
\end{equation}
\end{figure*}

\setcounter{equation}{\value{TempEqCnt}} % 把TempEqCnt中存的公式序号赋回给当前公式序号

The estimation of AoA relies on the linear phase shift on each antenna of the uniform linear array (ULA). However, the term $e^{j2\pi n\Delta f \tau _{\mathrm{STO},p}}$ related to the $p$-th antenna in Eq. (\ref{eq:errCSI}) will influence the linear phase shift represented by the steering vector $\boldsymbol{a}_{\boldsymbol{l}}[p]$.
In order to eliminate the impact of antenna-related STO $\tau _{\mathrm{STO},p}$, relative STO calibration is designed. With the raw CSI, we first conduct a $U$-point IFFT calculation to get the channel impulse response (CIR) in the time/delay domain for each antenna, given by

\begin{equation}
\begin{aligned}
&\hat{\mathbf{Q}}_{p}[m,u]= \frac{1}{U} \sum_{n=0}^{N-1} \hat{\mathbf{H}}_{p}[m,n] e^{j2\pi u n/U}\\
& =\frac{1}{U}\sum_{n=0}^{N-1} \sum_{l=1}^L{e^{j\theta _m}\boldsymbol{a}_{\boldsymbol{l}}[p]b_{l}}e^{j2\pi n\Delta f (\tau_{\mathrm{STO},p}-\tau _l)} e^{j\varphi_{\text{D},m,l}}e^{j2\pi u \frac{n}{U}},
\end{aligned}
\end{equation}
where a large $U$ can provide more accurate information for relative STO calibration.

Due to the different STO $\tau _{\mathrm{STO},p}$ in each port $p$, the CIR bin of the path varies in each port. However, since the propagation environment is assumed to be far-field, the same path should lie within the same CIR bin on every antenna. Based on this insight, the relevance of CIRs among each antenna is considered, and the relative STO can be estimated by finding the maximum relevance of CIRs. %The same experiment setup in the phase offset experiment can be reused here.
Particularly, the CIR from a certain antenna is fixed as the reference antenna. For CIRs from other antennas, they are slid circularly to find the best match with the CIR from the reference antenna. The best match represents the situation where the maximum sum of the product of the amplitudes of CIRs in the region of interest. To reduce computational complexity, only the bins surrounding significant peaks are considered in the relevance calculation. Specifically, the two largest local peaks are first identified. For each peak, the neighboring bins where the CIR power remained within 3 dB of the corresponding peak power are included in the region of interest. An example of relative STO calibration is shown in Fig. \ref{fig:relSTO}. The original CIRs from antenna 1 and antenna 2 are shown in Fig. \ref{fig:relSTO}(a). We fix the CIR from antenna 1 and slide the CIR from antenna 2, and the best match is shown in Fig. \ref{fig:relSTO}(b). The delay-overlapped CIR for antenna $p$, denoted as $\tilde{\mathbf{Q}}_{p}[m,u]$, is the relative-STO-calibrated CIR and the relative STO of antenna $p$ to antenna 1 can be represented as $\Delta_{\tau,p}=\tau_{\mathrm{STO,}p}-\tau_{\mathrm{STO,}1}$. The compensation value of the relative STO is thus obtained by $C_{p} = e^{j2\pi n\Delta f(\tau _{\mathrm{STO},p}-\tau _{\mathrm{STO},1})}$. With this value, the impact of relative STOs can be removed by
\begin{equation}
\begin{aligned}
\tilde{\mathbf{H}}_{p}[m,n] = &\hat{\mathbf{H}}_{p}[m,n]/C_{p}\\
 =& e^{j\theta _m}e^{j2\pi n\Delta f\tau _{\mathrm{STO},1}}\\
&\sum_{l=1}^L\boldsymbol{a}_{\boldsymbol{l}}[p]b_{l}e^{-j2\pi n\Delta f\tau _l}e^{j\varphi_{\text{D},m,l}}.
\end{aligned}
\label{eq:relSTO}
\end{equation}

After relative STO calibration, the processed CSI $\tilde{\mathbf{H}}_{p}$ can be obtained, which eliminates the terms related to antenna $p$, except for the steering vector. The distorted phase shift term $e^{j\theta _m}e^{j2\pi n\Delta f\tau _{\mathrm{STO},1}}$ is independent of antennas and can be considered as a common phase offset that will not influence AoA estimations, and therefore, the AoA estimation can be conducted. If we directly use $\tilde{\mathbf{H}}_{p}[m,n]$ to estimate AoAs of all paths along with antennas, we need a further step to match the delay of each path. To avoid this step, we directly extract the complex amplitude from several large local peaks on $\tilde{\mathbf{Q}}_{p}[m,u]$ and then estimate the AoAs for different delay taps.
%The processed CIR $\tilde{\mathbf{Q}}_{p}[m,u]$ is first obtained by applying the $U$ points IFFT to the processed CSI $\tilde{\mathbf{H}}_{p}[m,n]$.
Typical AoA estimation algorithms, such as MUSIC, are then utilized for AoA estimation based on the processed CIR in all antennas. The above steps leverage the delay-angle sparsity feature in mmWave systems.

All estimated AoAs of paths are compared with the AoA of the known reference path. %Since the channel of the millimeter wave system is sparse, we assume that all paths are unique in their incoming directions.
Therefore, the matched one is then identified as the reference path. With the above procedure, the reference path can be either the LoS path or the NLoS path from the reference reflector. If both exist, we can simply select the LoS path as the reference path.

%Since the location of the transmitter and receiver is known, the range and direction of the reference path can be calculated based on the geometry.
After identifying the reference path, we next remove the impact of $\theta _m$ and $\tau _{\mathrm{STO},1}$ with the given information of distance and AoA of the reference path.
The insights of the proposed calibration scheme lie in the fact that since the reference path and all other paths endure the same system errors, the distorted response of the reference path can be helpful in eliminating system errors in all paths. The channel matrix described in Eq. (\ref{eq:relSTO}) can be divided into two parts,
i.e., the reference path and non-reference paths, given by
\begin{equation}
\begin{aligned}
    \tilde{\mathbf{H}}_{p}[m,n]=&\underbrace{e^{j\theta _m}e^{j2\pi n\Delta f\tau _{\mathrm{STO},1}}}_{\text{remaining system errors}}
(\underbrace{\boldsymbol{a}_{\mathbf{ref}}[p]b_{\mathrm{ref}}e^{-j2\pi m\Delta f\tau _{\mathrm{ref}}}}_{\text{reference path}}\\
&+\underbrace{\sum_{\substack{l=1\\l\neq\mathrm{ref}}}^L{\boldsymbol{a}_{\boldsymbol{l}}[p]b_{l}}e^{-j2\pi m\Delta f\tau _l}e^{j\varphi_{\text{D},m,l}}}_{\text{non-reference paths}}).
\end{aligned}
\end{equation}
%where $e^{j\theta _m^\prime} = e^{j\theta _{\mathrm{offest},1}}e^{j\theta _m}$ is the product of the remaining phase offset and phase shift.
Note that since the reference path is static, there is no Doppler shift.
Due to the strong power of the reference path, we can directly extract the channel parameter of the reference path from $\tilde{\mathbf{Q}}_{p}[m,u_{\mathrm{r}}]$ at the delay bin of the reference path, which can be mathematically expressed as
%The channel parameter of the reference path can then be obtained from the CIR profile. The CIR of the reference path can be calculated with IFFT and extracted as
\begin{equation}
\begin{aligned}
\tilde{\mathbf{Q}}_{p}[m,u_{\mathrm{r}}] \approx & e^{j\theta _m}\boldsymbol{a}_{\mathbf{ref}}[p]b_{\mathrm{ref}},
\end{aligned}
\end{equation}
where $u_{\text{r}}$-th bin indicates the CIR index of reference path. We have the relationship of $\tau_{\mathrm{STO},1}=\tau_{\mathrm{ref}}-u_{\mathrm{r}}\delta$, where $\tau_{\text{ref}}$ is the true delay of the reference path and $\delta$ is the duration of a bin. Then, the STO can be eliminated by compensation, and the time-varying phase shift $\theta_m$ can be eliminated by a simple division, as given in Eq. (\ref{eq:CSIdiv}).

% \begin{equation}
%     \begin{aligned}
%         \hat{\mathbf{H}}_{p,\mathrm{div}}[m,n]
%         =&\tilde{\mathbf{H}}_p[m,n]e^{-j2\pi n\Delta f(\tau _{\mathrm{ref}}-u_{\mathrm{ref}}\delta )}/(\frac{\hat{\mathbf{Q}}_p[m,u_{\mathrm{ref}}]}{\left\| \hat{\mathbf{Q}}_p[m,u_{\mathrm{ref}}] \right\|})\\
%     = &\left[ e^{j\theta _m^\prime}e^{j2\pi n\Delta f\tau _{\mathrm{STO},1}}e^{-j2\pi n\Delta f(\tau _{\mathrm{ref}}-u_{\mathrm{ref}}\delta )} \right.\\
%     &\left( \boldsymbol{a}_{\mathbf{ref}}[p]b_{\mathrm{ref}}e^{-j2\pi n\Delta f\tau _{\mathrm{ref}}}+ \right. \\
%     & \sum_{\substack{l=1\\l\neq\mathrm{ref}}}^L{\boldsymbol{a}_{\boldsymbol{l}}[p]b_l}e^{-j2\pi n\Delta f\tau _l}e^{j\varphi_{\text{D},m,l}} )\\
%     & +\tilde{w}_{m,n} ] / \left( e^{j\theta _m^\prime}\boldsymbol{a}_{\mathbf{ref}}[p]\right)\\
%     =&b_{\mathrm{ref}}e^{-j2\pi m\Delta f\tau _{\mathrm{ref}}}\\
%     & +\sum_{\substack{l=1\\l\neq\mathrm{ref}}}^L{(\boldsymbol{a}_{\boldsymbol{l}}[p]/\boldsymbol{a}_{\mathbf{ref}}[p])b_l}e^{-j2\pi n\Delta f\tau _l}e^{j\varphi_{\text{D},m,l}})+\hat{w}_{\text{div},m,n}\\
%     \end{aligned}
%     \label{eq:CSIdiv}
% \end{equation}

However, after the division, the feature of the steering vector is destroyed in Eq. (\ref{eq:CSIdiv}), which can be reconstructed by multiplying the steering vector $\boldsymbol{a}_{\boldsymbol{\text{ref}}}[p]$ at the direction of the reference path. The final calibrated CSI is given by

\setcounter{equation}{8} % 当前公式序号变为y，y等于长公式的序号.
\begin{equation}
\begin{aligned}
    & \hat{\mathbf{H}}_{p,\mathrm{calib}}[m,n]=  \boldsymbol{a}_{\mathbf{ref}}[p]\hat{\mathbf{H}}_{p,\mathrm{div}}[m,n]\\ = &\boldsymbol{a}_{\mathbf{ref}}[p]b_{\mathrm{ref}}e^{-j2\pi n\Delta f\tau _{\mathrm{ref}}} +\sum_{\substack{l=1\\l\neq\mathrm{ref}}}^L{\boldsymbol{a}_{\boldsymbol{l}}[p]b_{l}}e^{-j2\pi n\Delta f\tau _l}e^{j\varphi_{\text{D},m,l}}.
\end{aligned}
\label{Eq:calibCSI}
\end{equation}

The calibrated CSI is now ready for further sensing processing to estimate the angle, delay, and Doppler shift of each path.

%After calibration is finished in each band, the calibrated CSI from all bands can be merged as $\hat{\mathbf{H}}_{p,\mathrm{calib}}$ for better sensing performance, given by
%\begin{equation}
%   \hat{\mathbf{H}}_{p,\mathrm{calib}}=[\hat{\mathbf{H}}_{1,p,\mathrm{calib}}^T, ..., \hat{\mathbf{H}}_{K,p,\mathrm{calib}}^T]^T.
%\end{equation}

%\input{performance}

\section{Performance Evaluation}

\subsection{Experimental Setup}
\begin{figure}[t]
    \centering
    \includegraphics[width=0.7\linewidth]{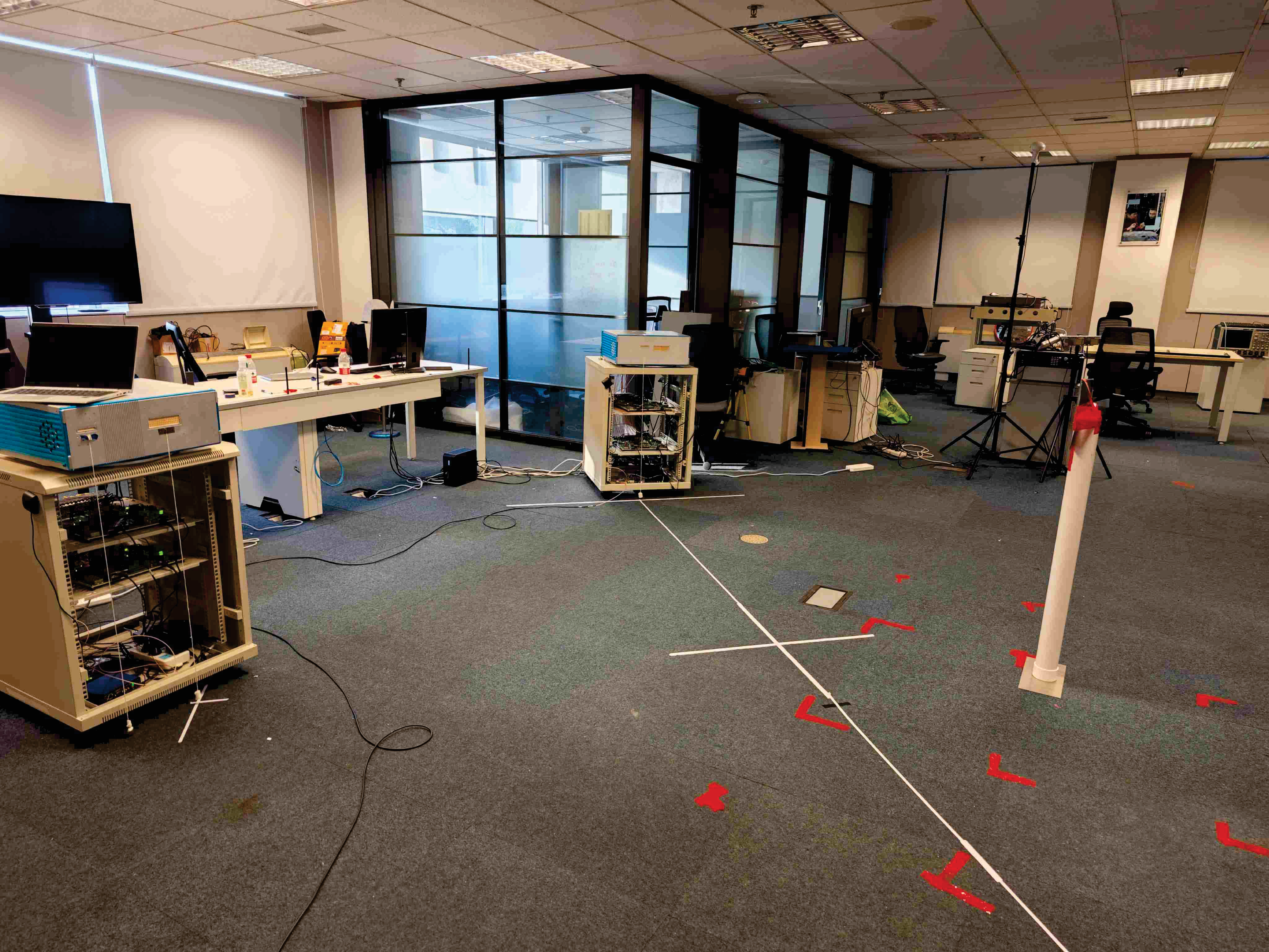}
    \caption{Experiment environment.}
    \vspace{-1em}
    \label{fig:ESsetup}
\end{figure}

Experiments are conducted to validate the performance of the proposed system calibration scheme. We develop a 26 GHz mmWave ISAC prototype system. Particularly, the system implements the OFDM waveform with 1824 subcarriers with a carrier spacing of 270 kHz and a total bandwidth of 500 MHz. Within the 1824 subcarriers, one every 24 subcarriers is placed a reference signal (RS). A total of $N=76$ subcarriers are utilized as RS subcarriers, and the frequency interval of the RS subcarriers is 6.48 MHz. The duration of an OFDM block is 4 ms, which consists of 864 symbols, each symbol with 4.67 $\mu$s of duration. An RS symbol is deployed once every OFDM block, which means the time interval between two adjacent RS symbols is 4 ms. Ten adjacent RS symbols are concatenated together for a round of the sensing process, which is denoted as a sensing frame. %Both the transmitter and the receiver utilized the same set of hardware.
At the transmitter side, one transmit antenna is utilized to transmit the OFDM signal. The transmission power is 0 dBm. At the receiver side, a uniform linear array of 1$\times$8 antennas is utilized for receiving the OFDM signal, with the antenna interval of half wavelength $\Delta_d = \lambda/2 = 5.7$ mm. Each receive antenna is connected to an RF chain independently. Communication synchronization techniques are applied to get achieve the OFDM receiving synchronization. The experiments are conducted in an indoor office environment, as shown in Fig. \ref{fig:ESsetup}.

\subsection{Experimental Results}

\begin{figure}[t]
    \centering
	\includegraphics[width=0.9\linewidth]{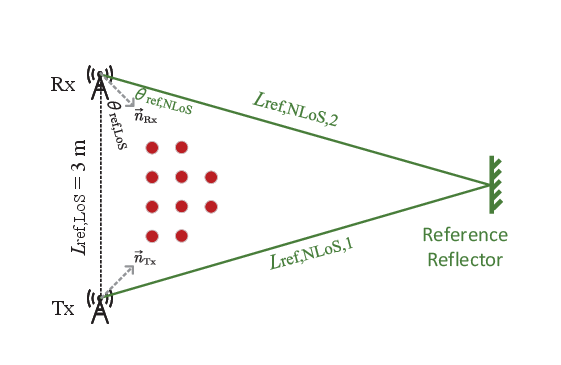}
    \vspace{-2em}
    \caption{Experiment setup for validation of reference-path-aided calibration.}
    \vspace{-1em}
    \label{fig:ES_setupRefCalib}
\end{figure}

The performance of the reference-path-aided calibration is verified in the cases that the LoS path or the NLoS path is utilized as the reference path, respectively. In the experiment, we assume that the position of Tx and Rx is known. Consider the setup with coordinates of the room shown in Fig. \ref{fig:ES_setupRefCalib}, where the Tx is fixed at (0, 0) m and the Rx is fixed at (-3, 0) m. The orientations of Tx and Rx are fixed, whose normal direction vector is set as $\vec{n}_{\text{Tx}} = (-\sqrt{2}/2,\sqrt{2}/2)$ and $\vec{n}_{\text{Rx}} = (\sqrt{2}/2,\sqrt{2}/2)$, respectively. The range of the LoS path is given by $L_{\text{ref, LoS}} = 3$ m and the AoA of the LoS path is $\theta_{\text{ref, LoS}} = -45^\circ$. A metal plate is placed on the desk far away as the NLoS reference reflector in the NLoS case, whose central point is at (-1.5, 13.3) m, and its height is the same as both Tx and Rx. The physical range of the NLoS reference path of Tx-plate-Rx is measured as $L_{\text{ref, NLoS}} = L_{\text{ref, NLoS},1}+L_{\text{ref, NLoS},2} = 26.77$ m, and the AoA of the NLoS reference path is measured as $\theta_{\text{ref, NLoS}} = 38.5^\circ$. A curved plastic sheet of the size of 30 cm $\times$ 22 cm is mounted on a base as the static target for localization performance evaluation, which is placed in multiple places as marked by the red dots in Fig. \ref{fig:ES_setupRefCalib}. For the evaluation of the NLoS case, we manually block the LoS path. From our setup, we can see that the delay of the target is less than the delay of the reference NLoS path.

\begin{figure}[t!]
    \centering
    \includegraphics[width=0.7\linewidth]{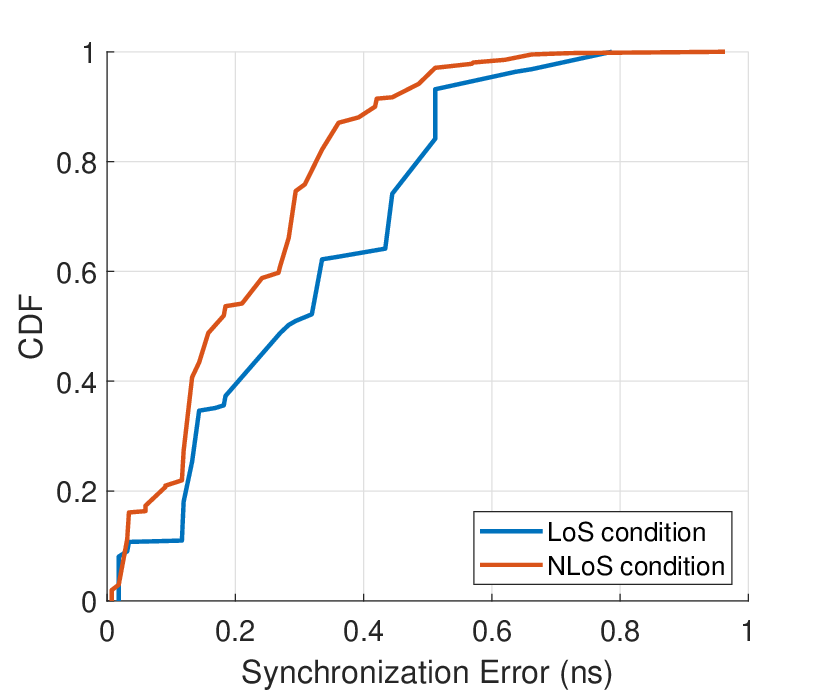}
    \caption{CDF of the synchronization error in LoS and NLoS conditions.}
    \vspace{-1em}
    \label{fig:RefCalibsyn}
\end{figure}

\begin{figure}[t!]
    \centering
    \includegraphics[width=0.7\linewidth]{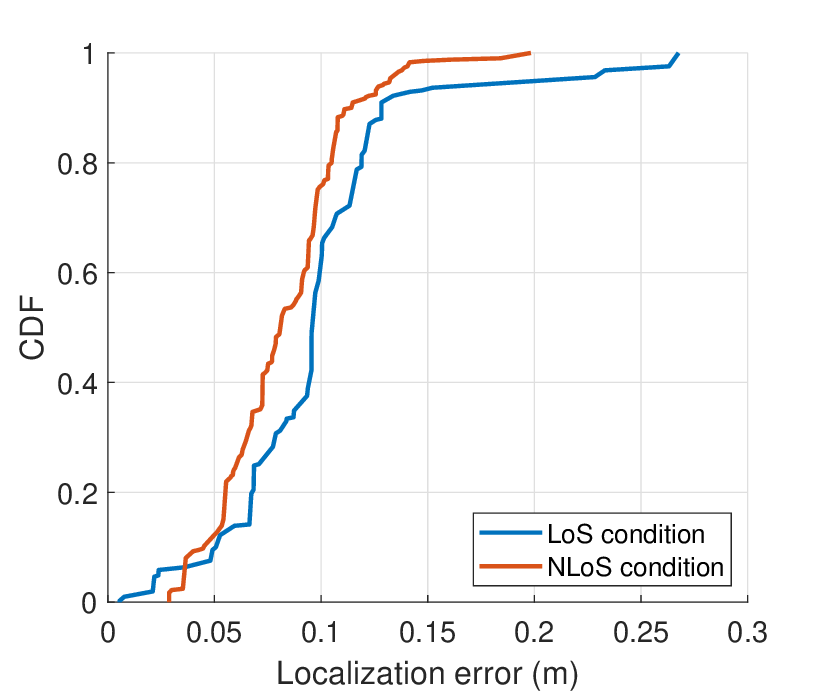}
    \caption{CDF of the localization error in LoS and NLoS conditions.}
    \vspace{-1em}
    \label{fig:RefCalibLoc}
\end{figure}

During the validation, our proposed reference-path-aided calibration is first conducted. Next, the target location is estimated via the estimated delay and AoA information. The delay synchronization error is calculated by considering only the range estimation error. The target localization error is also calculated, which further includes the influence of angle estimation error.

The delay synchronization performance of our reference-path-aided calibration in both the LoS and NLoS scenarios is shown in Fig. \ref{fig:RefCalibsyn}. In the LoS condition, the average synchronization error is 0.3 ns, while the 80th percentile error is 0.4 ns, and the maximum synchronization error is less than 0.8 ns. In the NLoS condition, the average synchronization error is 0.2 ns, while the 80th percentile error is 0.3 ns, and the maximum synchronization error is less than 1 ns. Therefore, the synchronization error in both cases is less than 1 ns, which also satisfies the maximum tolerance of error in 3GPP TS 38.141 \cite{38141}. The localization error in both conditions is illustrated in Fig. \ref{fig:RefCalibLoc}. In the LoS condition, the average localization error is 10 cm, while the 80th percentile error is 12 cm, and the maximum localization error is 26 cm. In the NLoS condition, the average localization error is 8 cm, while the 80th percentile error is 10 cm, and the maximum localization error is 19 cm.
%For the delay synchronization and localization errors, the LoS case and NLoS case have a similar performance, or even the NLoS case has a little bit better performance.
For delay synchronization and localization errors, the LoS and NLoS cases exhibit similar performance, with the NLoS case even performing slightly better.
This slight difference may come from the ground truth we used. Our sensing target is a plastic sheet with a width of 30 cm. The central point of the sheet is utilized as the reflection point for ground truth. In practice, the signal can be reflected from any point on the sheet instead of exactly the central point of the sheet. Therefore, an average localization error of around 10 cm is within expectation, and such a ground truth selection may cause a bias, resulting in a slightly better performance of the NLoS case.
These experiments demonstrate that our proposed reference-path-aided calibration achieves performance sufficient to support high-accuracy range-related sensing applications.

\begin{figure}[t!]
    \centering
    \subfigure[]{
        \includegraphics[width=0.46\linewidth]{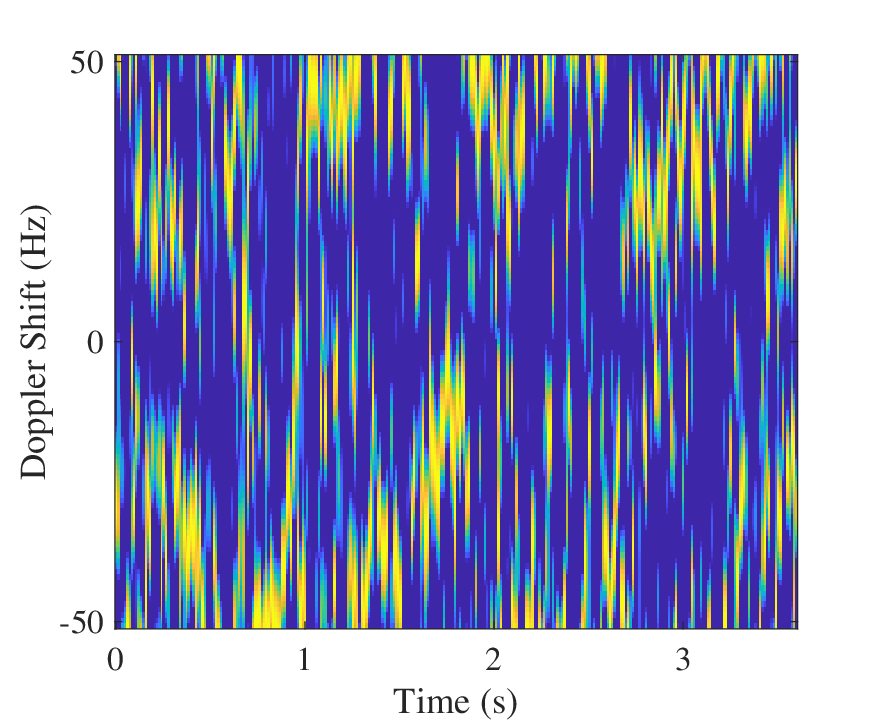}
        \label{fig:DSraw}
    }
    \subfigure[]{
	    \includegraphics[width=0.46\linewidth]{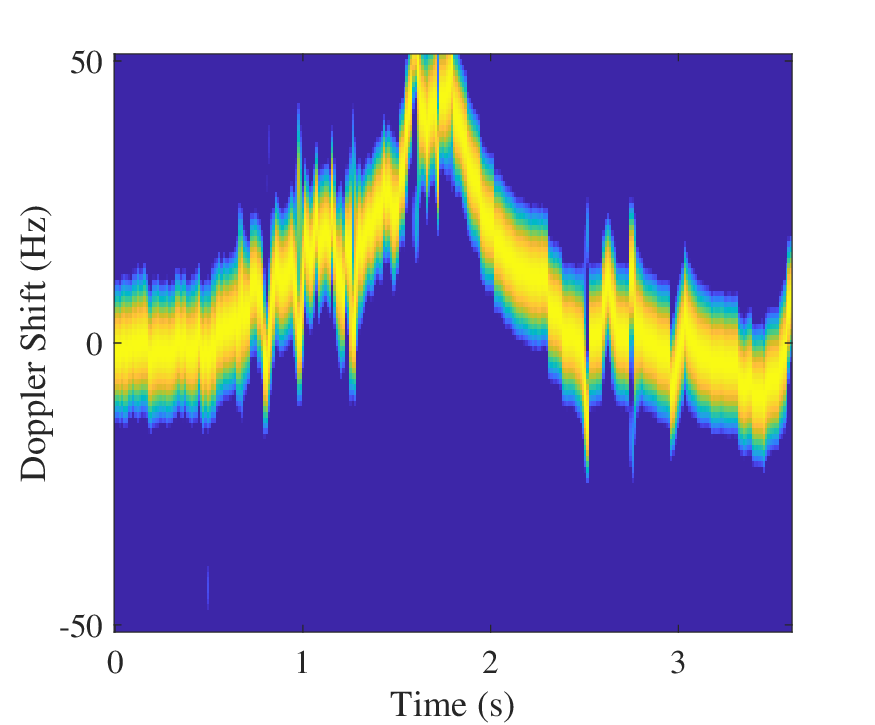}
        \label{fig:DScalib}
    }
    \caption{Doppler spectrum of a person standing up: (a) before calibration, the pattern of motion is affected by the random phase shift; (b) the random phase shift is eliminated by calibration.}
    \vspace{-2em}
    \label{fig:DS}
\end{figure}

Next, the effectiveness of utilizing the proposed calibration scheme in extracting the Doppler information is demonstrated. More specifically, a person is standing up in this motion. The Doppler spectrum of the human target is then extracted, as shown in Fig. \ref{fig:DS}. Due to the time-varying phase change, especially the random phase shift, the uncalibrated Doppler spectrum appeared distorted, and the motion pattern cannot be identified, as shown in Fig. \ref{fig:DSraw}. After applying the reference-path-aided calibration, the Doppler pattern becomes significantly more distinguishable, as illustrated in Fig. \ref{fig:DScalib}. %This Doppler result matches the ground-truth motion pattern measured by a radar system, as reported in \cite{act}.
This result demonstrates that the proposed calibration method can effectively mitigate the time-varying system error and enable accurate Doppler estimation for Doppler-related sensing applications.

\section{Conclusion}

In this paper, a reference-path-aided system calibration scheme was proposed for mmWave bistatic ISAC systems. The proposed approach utilized a reference path in the environment to facilitate accurate calibration, which was feasible in both LoS and NLoS propagation conditions. By exploiting the consistency of channel response of the reference path, the system compensated for impairment errors, including STO, CFO, and random phase shifts. An mmWave ISAC prototype system was developed to validate the effectiveness of our proposed scheme. Experimental results showed that a delay synchronization error within 1 ns can be achieved, and the time-varying errors can be effectively eliminated, thus meeting the requirements of high-precision ISAC applications.

%\section{Acknowledgement}
%The research work is supported by National Natural Science Foundation of China (NSFC), No. 61801290.


\bibliographystyle{IEEEtran}
\bibliography{reference}

\end{document}